\documentclass[english,aps,twocolumn,prl,superscriptaddress,showpacs]{revtex4}
\usepackage[T1]{fontenc}
\usepackage[latin9]{inputenc}
\usepackage{amsmath}
\usepackage{graphicx}
\usepackage{amsthm,amsfonts,amssymb,times,bbm}

\def\RR{\mathbbm{R}}
\newcommand{\dicke}[2]{\ket{D_{#1}^{(#2)}}\,}
\newcommand{\ket}[1]{\mbox{$\,\mid \! #1 \, \rangle$}}

\newcommand{\citeappendix}[1]{}
\newcommand{\citepreprint}[1]{{#1}}



\usepackage{babel}

\begin{document}

\title{Permutationally Invariant Quantum Tomography}

\date{\today}

\author{G\'eza~T\'oth}
\affiliation{Department of Theoretical Physics, The University of the Basque Country,
P.O. Box 644, E-48080 Bilbao, Spain}
\affiliation{IKERBASQUE, Basque Foundation for Science, E-48011 Bilbao, Spain}
\affiliation{Research Institute for Solid State Physics and Optics, Hungarian
Academy of Sciences, P.O. Box 49, H-1525 Budapest, Hungary}
\author{Witlef~Wieczorek}
\altaffiliation{Present address: Faculty of Physics, University of Vienna,
Boltzmanngasse 5, A-1090 Wien, Austria}
\affiliation{Max-Planck-Institut f\"ur Quantenoptik, Hans-Kopfermann-Strasse
1, D-85748 Garching, Germany}
\affiliation{Fakult\"at f\"ur Physik, Ludwig-Maximilians-Universit\"at, D-80797 M\"unchen, Germany}
\author{David~Gross}
\affiliation{Institute for Theoretical Physics, Leibniz University Hannover, D-30167
Hannover, Germany}
\author{Roland~Krischek}
\affiliation{Max-Planck-Institut f\"ur Quantenoptik, Hans-Kopfermann-Strasse
1, D-85748 Garching, Germany}
\affiliation{Fakult\"at f\"ur Physik, Ludwig-Maximilians-Universit\"at, D-80797
M\"unchen, Germany}
\author{Christian~Schwemmer}
\affiliation{Max-Planck-Institut f\"ur Quantenoptik, Hans-Kopfermann-Strasse
1, D-85748 Garching, Germany}
\affiliation{Fakult\"at f\"ur Physik, Ludwig-Maximilians-Universit\"at, D-80797
M\"unchen, Germany}
\author{Harald~Weinfurter}
\affiliation{Max-Planck-Institut f\"ur Quantenoptik, Hans-Kopfermann-Strasse
1, D-85748 Garching, Germany}
\affiliation{Fakult\"at f\"ur Physik, Ludwig-Maximilians-Universit\"at, D-80797
M\"unchen, Germany}

%
%
%
%
%
%


\begin{abstract}
We present a scalable method for the tomography of large multiqubit
quantum registers. It acquires information about the
permutationally invariant part of the density operator, which is a
good approximation to the true state in many, relevant cases.
Our method gives the best measurement strategy to minimize
the experimental effort as well as the uncertainties of the reconstructed density matrix.
We apply our method to the experimental tomography of a
photonic four-qubit symmetric Dicke state.
\end{abstract}
\maketitle

Because of the the rapid development of quantum experiments, it
is now possible to create highly entangled multiqubit states using photons
\cite{PB00,KS07,WK09,W08,PC09}, trapped ions
\cite{SK00}, and cold atoms \cite{MG03}. So far, the largest
implementations that allow for an individual readout of the particles
involve on the order of $10$ qubits. This number will soon be overcome, for
example, by using several degrees of freedom within each particle to
store quantum information \cite{CB05}.
Thus, a new regime will be reached in
which a complete state tomography is impossible even from the point of
view of the storage place needed on a classical computer. At this point the
question arises: Can we still extract useful information about the
quantum state created?

In this Letter we propose permutationally invariant (PI) tomography in multiqubit
quantum experiments \cite{Other}.
Concretely, instead of the density matrix $\varrho,$ we propose to
determine the PI part of the density matrix
defined as
\begin{equation}
\varrho_{{\rm PI}}=\frac{1}{N!}\sum_{k}\Pi_{k}\varrho\Pi_{k},\label{rhopi}\end{equation}
where $\Pi_{k}$ are all the permutations of the qubits.
Reconstructing $\varrho_{{\rm PI}}$ has been considered theoretically
for spin systems (see, e.g., Ref.~\cite{AM03}).
Recently it has been pointed out that
photons in a single mode optical fiber will always be in a PI state
and that there is only a small set of measurements needed for their characterization \cite{AS07,SHALM09}.

Here, we develop a provably optimal scheme,
which is feasible for large multiqubit systems:
For our method, the measurement effort increases only
\emph{quadratically} with the size of the system.
Our approach is further motivated by the fact that
 almost all multipartite 
experiments are done with PI quantum states \cite{SK00,PB00,KS07,WK09,PC09}. Thus, the density matrix
obtained from PI tomography is expected to be
close to the one of the experimentally achieved state.
The expectation values of symmetric operators, such as some entanglement witnesses, and fidelities
with respect to symmetric states
are the same for both density matrices and are thus obtained exactly from
PI tomography \cite{KS07,WK09,PC09}.
Finally, if $\varrho_{\rm PI}$ is entangled, so is the state $\varrho$ of the system,
which makes PI tomography a useful and efficient tool for
entanglement detection.

Below, we summarize the four main contributions of this Letter.
We restrict our attention to the case of $N$ qubits --
higher-dimensional systems can be treated similarly.

1.\ In most experiments, the qubits can be individually
addressed whereas nonlocal quantities cannot be measured directly.  The
experimental effort is then 
characterized by the number of
local measurement settings needed,
where ``setting'' refers to the choice of
one observable per qubit, and repeated von Neumann measurements in the
observables' eigenbases \cite{Remark}. Here, we compute the minimal number
of measurement settings required to recover $\varrho_{\rm PI}.$

2.\ The requirement that the number of settings be minimal does not
uniquely specify the tomographic protocol. On the one hand,
there are infinitely many possible choices for the local 
settings that are both minimal and give sufficient information to
find $\varrho_{\rm PI}$. On the other hand, for each given setting, there
are many ways of estimating the unknown density operator from the
collected data.
We present a systematic method to find the optimal scheme
through statistical error analysis.

3.\ Next, we turn to the important problem of gauging the information
loss incurred due to restricting attention to the PI
part of the density matrix. We describe an easy test measurement
that can be used to judge the applicability of PI\
tomography
\emph{before} it is
implemented.

4.\ Finally, we demonstrate that these techniques are viable in practice by
applying them to a photonic experiment observing a four-qubit
symmetric Dicke state.

\textbf{\emph{Minimizing the number of settings.}}
We will now present our first main result.\\
\textbf{\emph{Observation
1.}} For a system of $N$ qubits, permutationally invariant tomography
can be performed with
\begin{equation}
	\mathcal{D}_{N} ={{N+2}\choose{N}} =\frac12(N^2+3N+2)
	\label{eq:NDfull}
\end{equation}
local settings. It is not possible to perform such a tomography with fewer settings.\\
{\it Proof.}
First, we need to understand the information obtainable from a single
measurement setting. We assume that for every given setting, the same
basis is measured at every site \cite{AAA}.
Measuring a local basis $\{|\phi_1\rangle, |\phi_2\rangle\}$ is
equivalent to estimating the expectation value of the traceless
operator
$A=|\phi_1\rangle\langle\phi_1|-|\phi_2\rangle\langle\phi_2|$.
Merely by measuring $A^{\otimes N},$
it is possible to obtain all the $N$ expectation values
\begin{equation}\label{eq:symExpectations}
	\langle (A^{\otimes (N-n)}\otimes\openone^{\otimes n})_{\rm
	PI}\rangle,\qquad
(n=0, \dots, N-1)
\end{equation}
and, conversely, that is all the information obtainable about $\varrho_{\rm PI}$
from a single setting.

Next, we will use the fact that any PI density operator can be written as a linear
combination of the pairwise orthogonal operators $(X^{\otimes
k}\otimes Y^{\otimes l}\otimes Z^{\otimes m}\otimes\openone^{\otimes
n})_{{\rm PI}},$ where $X,Y$ and $Z$ are the Pauli matrices.
We consider the space spanned by these operators for one specific value
of $n$. Simple counting shows that its dimension is
$\mathcal{D}_{(N-n)}.$ The same space is spanned by
$\mathcal{D}_{(N-n)}$ generic operators of the type $(A^{\otimes
(N-n)}\otimes\openone^{\otimes n})_{\rm PI}.$ We draw two
conclusions: First, any setting gives at most one expectation value for every
such space.  Hence the number of settings cannot be smaller than the
largest dimension, which is $\mathcal{D}_N$. Second, a generic choice
of $\mathcal{D}_N$ settings is sufficient to recover the correlations
in each of these spaces, and hence completely characterizes $\varrho_{\rm
PI}$. This concludes the proof \cite{Remark_Obs1}.

The proof implies that there are real
coefficients $c^{(k,l,m)}_j$ such that
\begin{eqnarray}
 &  & \langle(X^{\otimes k}\otimes Y^{\otimes l}\otimes Z^{\otimes m}\otimes\openone^{\otimes n})_{{\rm PI}}\rangle=\nonumber \\
 &  & \;\;\;\;\;\;\;\;\;\;\;\;\;\;\;\sum_{j=1}^{\mathcal{D}_{N}} c_{j}^{(k,l,m)}\langle(A_{j}^{\otimes(N-n)}\otimes
 \openone^{\otimes n})_{{\rm PI}}\rangle.\label{eq:XYZ2}
\end{eqnarray}
We will refer to the numbers on the left-hand side of Eq.~(\ref{eq:XYZ2}) as the elements of the
\emph{generalized Bloch vector}. The expectation values on the right-hand side can be
obtained by measuring the settings with $A_{j}$ for
$j=1,2,...,\mathcal{D}_{N}.$

\textbf{\emph{Minimizing uncertainties.}}
We now have to determine the optimal scheme for PI tomography.
To this end, we define our measure of statistical uncertainty as
the sum of the
variances of all the Bloch vector elements
\begin{eqnarray}
 &  & (\mathcal{E}_{{\rm total}})^{2}=\nonumber \\
 &  & \sum_{k+l+m+n=N}\mathcal{E}^{2}\left[(X^{\otimes k}\otimes Y^{\otimes l}\otimes
 Z^{\otimes m}\otimes\openone^{\otimes n})_{{\rm PI}}\right]\times\nonumber \\
 &  & \left(\frac{N!}{k!l!m!n!}\right),\label{eq:ET}\end{eqnarray}
where the term with the factorials is the number of different permutations
of $X^{\otimes k}\otimes Y^{\otimes l}\otimes Z^{\otimes m}\otimes\openone^{\otimes n}$.
Based on Eq.~\eqref{eq:XYZ2}, the variance of a single Bloch vector element is
\begin{eqnarray}
	&&
	\mathcal{E}^{2}[(X^{\otimes k}\otimes Y^{\otimes l}\otimes
	Z^{\otimes m}\otimes\openone^{\otimes n})_{{\rm PI}}] \nonumber\\
	&&\;\;\;\;\;\;\;\;\;=
	\sum_{j=1}^{\mathcal{D}_{N}}\vert
	c_{j}^{(k,l,m)}\vert^{2}\mathcal{E}^{2}[(A_{j}^{\otimes(N-n)}\otimes\openone^{\otimes
	n})_{{\rm PI}}].\label{eq:XYZ}
\end{eqnarray}
Equation (\ref{eq:ET}) can be minimized by changing the $A_j$ matrices and the $c_j^{(k,l,m)}$ coefficients.
We consider the coefficients first.
For any Bloch vector element, finding $c_j^{(k,l,m)}$'s that  minimize
the variance Eq.~(\ref{eq:XYZ})
subject to the constraint that equality holds in Eq.~(\ref{eq:XYZ2})
is a least squares problem. It has an analytic solution
obtained as follows: Write the operator on the left-hand side
of Eq.~(\ref{eq:XYZ}) as a vector $\vec{v}$ (with respect to some basis). 
Likewise, write the operators on the right-hand side
as $\vec{v}_j$ and define
a matrix $V=[\vec{v}_1, \vec{v}_2, \dots, \vec{v}_{\mathcal{D}_N}]$. Then Eq.~(\ref{eq:XYZ2})
can be cast into the form
$\vec{v}=V\vec{c}$, where $\vec{c}$ is a vector of the $c_j^{(k,l,m)}$ values for given $(k,l,m).$ If $E$ is the diagonal matrix
with entries
$E_{j,j}^2=\mathcal{E}^{2}[(A_{j}^{\otimes(N-n)}\otimes\openone^{\otimes
n})_{{\rm PI}}]$, then the optimal solution is
$\vec{c}=E^{-2} V^T (V E^{-2} V^T)^{-1} \vec{v},$
where the inverse is taken over the range \citeappendix{(see the Appendix for a proof).}
\citepreprint{\cite{preprint}.}

Equipped with a method for obtaining the optimal $c_j^{(k,l,m)}$'s for
 every fixed set of observables $A_j,$
it remains to find the best settings to measure. Every qubit
observable can be defined by the measurement directions
$\vec{a}_j$ using $A_j=a_{j,x}X+a_{j,y}Y+a_{j,z}Z.$ Thus, the task is
to identify $\mathcal{D}_N$ measurement directions on the
Bloch sphere minimizing the variance. In general, finding the globally optimal
solution of high-dimensional problems is difficult.
In our case, however, $\mathcal{E}_{\rm total}$ seems to penalize an
inhomogeneous distribution of the $\vec{a}_j$ vectors;
thus, using evenly distributed vectors as an initial guess, 
usual minimization procedures can be used to decrease $\mathcal{E}_{\rm total}$
and obtain satisfactory results \cite{preprint}.

The variance
$\mathcal{E}^{2}[(A_{j}^{\otimes(N-n)}\otimes\openone^{\otimes
n})_{{\rm PI}}]$ of the observed quantities depends on the physical
implementation. In the photonic setup below, we assume Poissonian
distributed counts.
It follows that (see also Refs.~\cite{S09,JN09})
\begin{equation}
	\mathcal{E}^{2}[
(A_{j}^{\otimes(N-n)}\otimes\openone^{\otimes n})_{{\rm PI}}
]
 =\frac{
	 [\Delta
	 (A_{j}^{\otimes(N-n)}\otimes\openone^{\otimes n})_{{\rm
	 PI}}]^2_{\varrho_0}
 }
 {\lambda_j-1},\label{eq:dddd}
\end{equation}
where  $(\Delta A)^2_\varrho=\langle A^2\rangle_{\varrho}-\langle
A\rangle_{\varrho}^2,$ $\varrho_{{\rm 0}}$ is the state of the system
and $\lambda_j$ is the parameter of the Poissonian distribution, which
equals the expected value of the total number of counts for the
setting $j.$
The variance
depends on the unknown state.
If we have preliminary knowledge of the likely form of $\varrho_0$,
we should use that information in the optimization.
Otherwise, $\varrho_0$ can be set to the completely mixed state.
For the latter, straightforward calculation shows that
$\mathcal{E}^{2}[ (A_{j}^{\otimes(N-n)}\otimes\openone^{\otimes n})_{{\rm PI}} ]=\binom{N}{n}^{-1} /(\lambda_j-1).$
For another implementation, such as trapped ions,
our scheme for PI tomography can be used after replacing Eq.~(\ref{eq:dddd}) by
a formula giving the variance for that implementation.

\textbf{\emph{Estimating the information loss due to symmetrization.
}}It is important to know how close the PI quantum state is
to the state of the system as PI tomography should serve as an alternative
of full state tomography for experiments aiming at the preparation of PI states.
\\
\textbf{Observation 2. }The fidelity between the original state
and the permutationally invariant state, $F(\varrho,\varrho_{{\rm PI}}),$ can be estimated from below as
$F(\varrho,\varrho_{{\rm PI}}) \ge\langle P_{{\rm s}}\rangle_{\varrho}^{2},$
 where $P_{{\rm s}}=\sum_{n=0}^{N}\vert D_{N}^{(n)}\rangle\langle D_{N}^{(n)}\vert$ is the projector
 to the $N$-qubit symmetric subspace, and the symmetric Dicke state is defined as
 $
\vert D_{N}^{(n)}\rangle=\binom{N}{n}^{-\frac{1}{2}}\sum_{k}\mathcal{P}_{k}(\vert0\rangle^{\otimes(N-n)}\otimes\vert1\rangle^{\otimes n}),$ 
where the summation is over all the different permutations of the
qubits.
Observation 2 can be proved based on Ref.~\cite{Subfidelity}
 and elementary matrix manipulations. Note that Observation 2
makes it possible to estimate $F(\varrho,\varrho_{{\rm PI}})$ based
on knowing only $\varrho_{{\rm PI}}.$

Lower bounds
on the fidelity to symmetric Dicke states, i.e., ${\rm Tr}(\vert D_{N}^{(n)}\rangle\langle D_{N}^{(n)}\vert\varrho)$
can efficiently be obtained by measuring $X,$ $Y$ and  $Z$ on all qubits,
i.e., measuring only three local settings  independent of $N$ \cite{TW09} .
With the same measurements,
one can also obtain a lower bound on the overlap between the state and the symmetric subspace.
For four qubits, this can be done based on
$P_{\rm s}\ge [(J_x^4+J_y^4+J_z^4)
-(J_x^2+J_y^2+J_z^2)]/18,$ where $J_x=(1/2)\sum_k X_k,$ $J_y=(1/2)\sum_k Y_k,$ etc.
Operators for estimating $\langle P_{\rm s}\rangle$ for $N=6,8$ are given
\citeappendix{in the Appendix.}\citepreprint{in Ref.~\cite{preprint}.}
 This allows one to judge how suitable the quantum state
is for PI tomography {\it before} such a tomography is carried out.

\begin{figure}
\begin{centering}
\includegraphics[width=5.2cm]{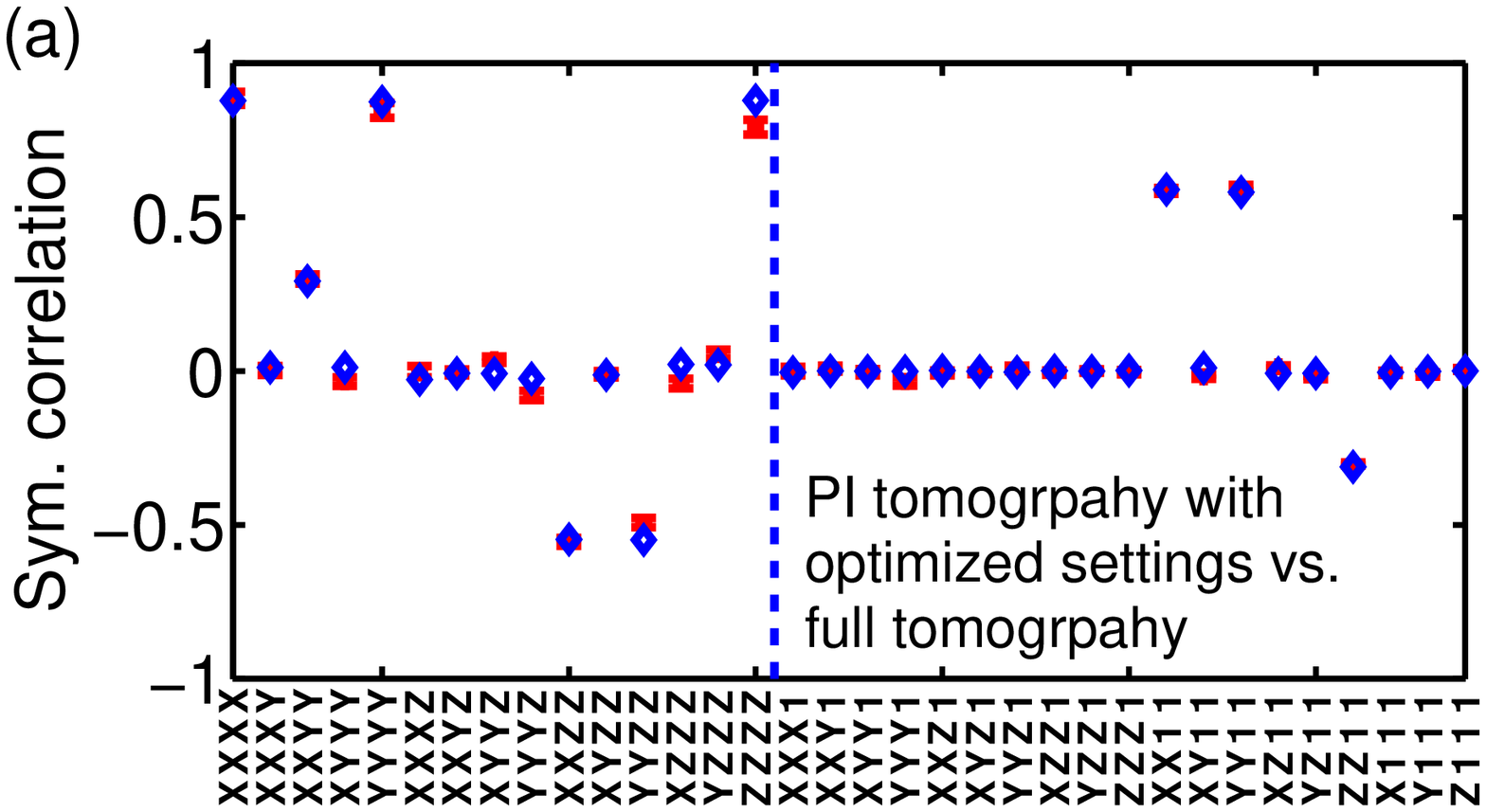}\includegraphics[width=3.3cm]{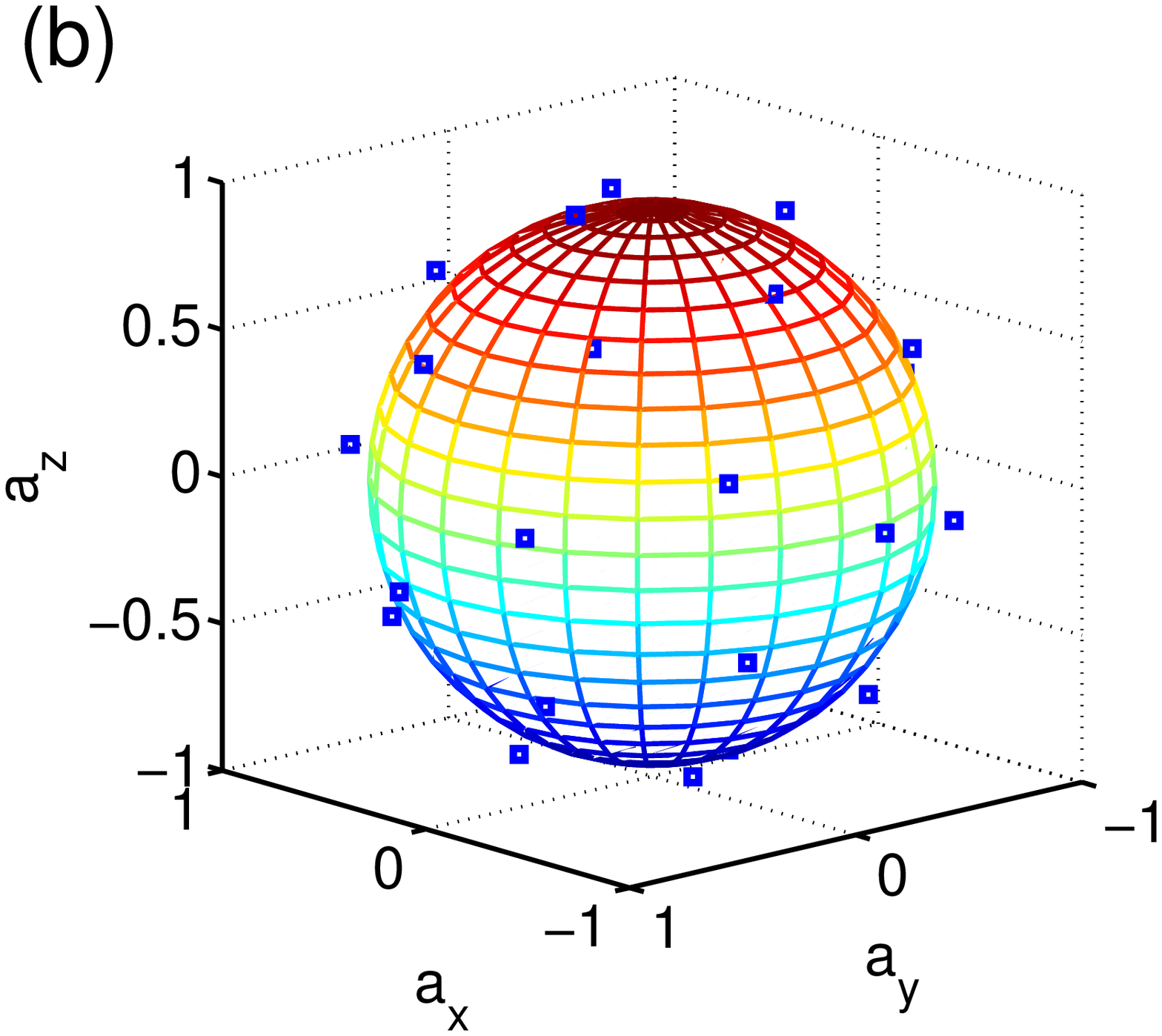}
\par\end{centering}
\begin{centering}
\includegraphics[width=5.2cm]{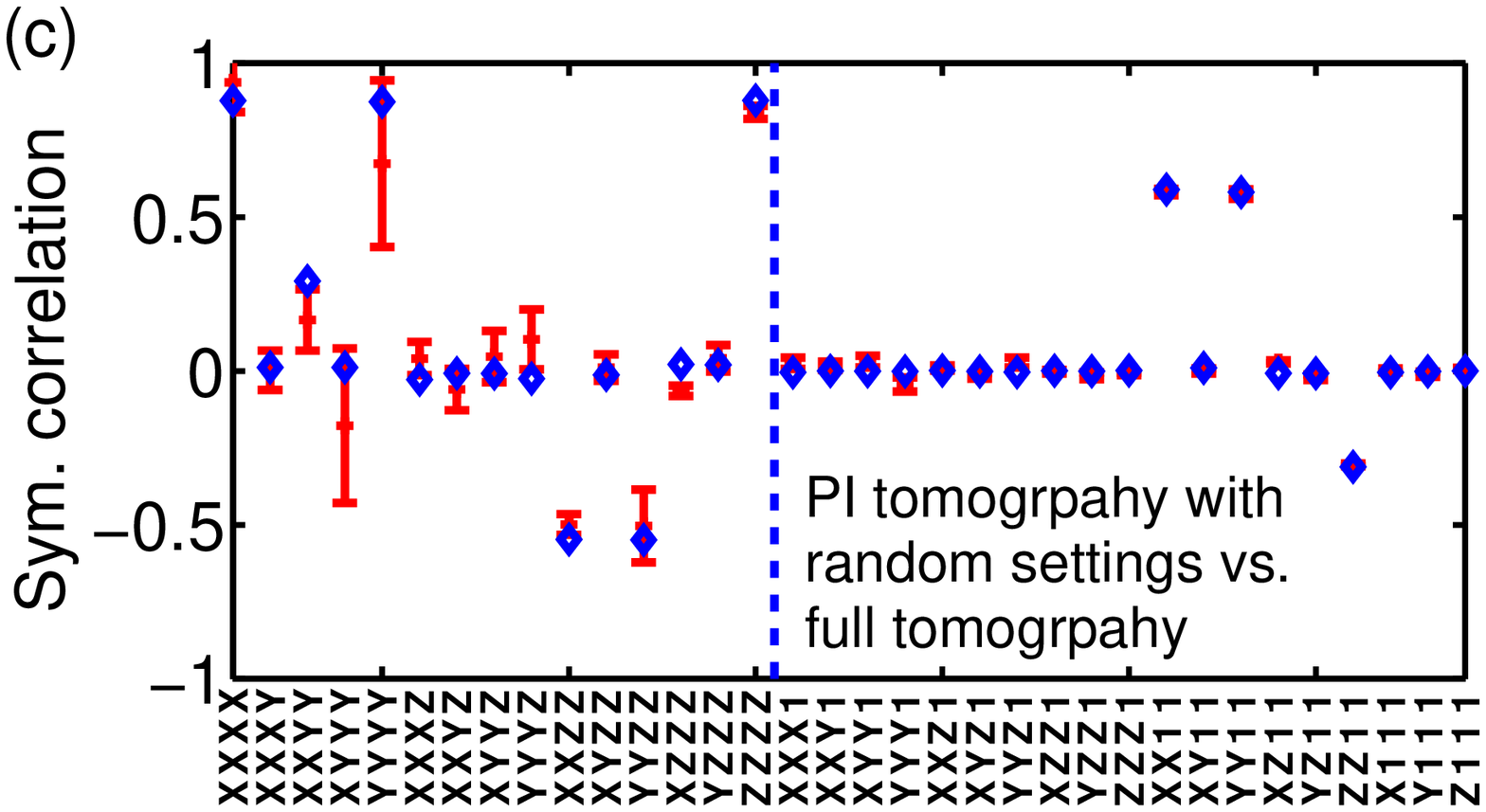}\includegraphics[width=3.3cm]{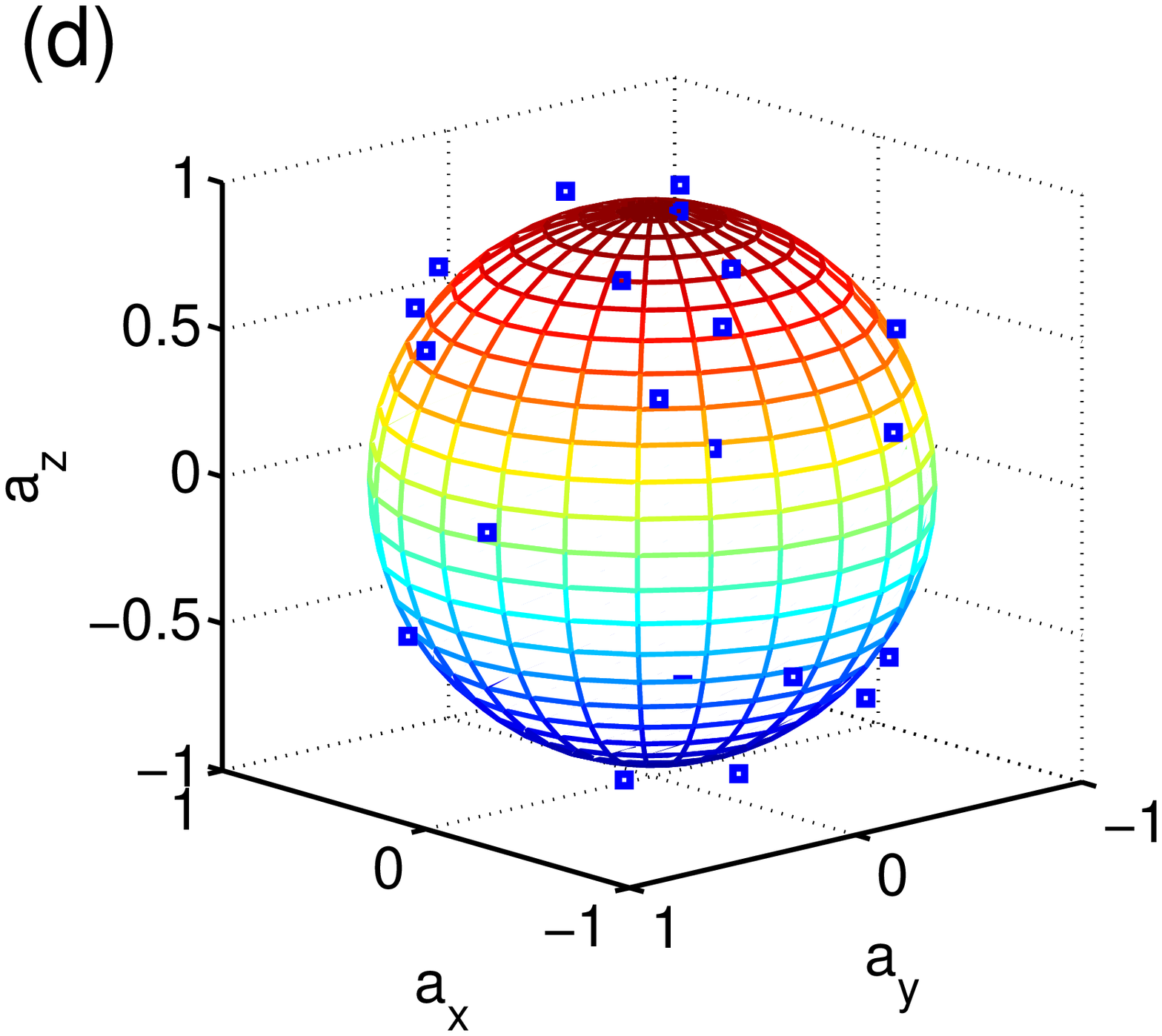}
\par\end{centering}
\caption{\label{fig:Error-of-the}(Color online) (a) Comparison of the $34$ symmetrized
correlations coming from (crosses with error bars) $15$ permutationally invariant measurement settings
with optimized $A_j$ matrices for $N=4$ qubits and (diamonds) from full tomography requiring $81$ local settings.
The average uncertainty
of all symmetrized correlations obtained from full tomography is
$\pm0.022$, and is not shown in the figure.
The labels refer to symmetrized correlations of the form given in the left-hand side of Eq.~(\ref{eq:XYZ2}).
 The results corresponding to the $15$ full four-qubit
correlations are left from the vertical dashed line. (b) Measurement
directions. A point at $(a_{x},a_{y},a_{z})$ corresponds to measuring
operator $a_{x}X+a_{y}Y+a_{z}Z.$ (c) Results for randomly chosen
$A_j$ matrices and (d) corresponding measurement directions.}
\end{figure}

\begin{figure}
\begin{centering}
\includegraphics[width=7.1cm]{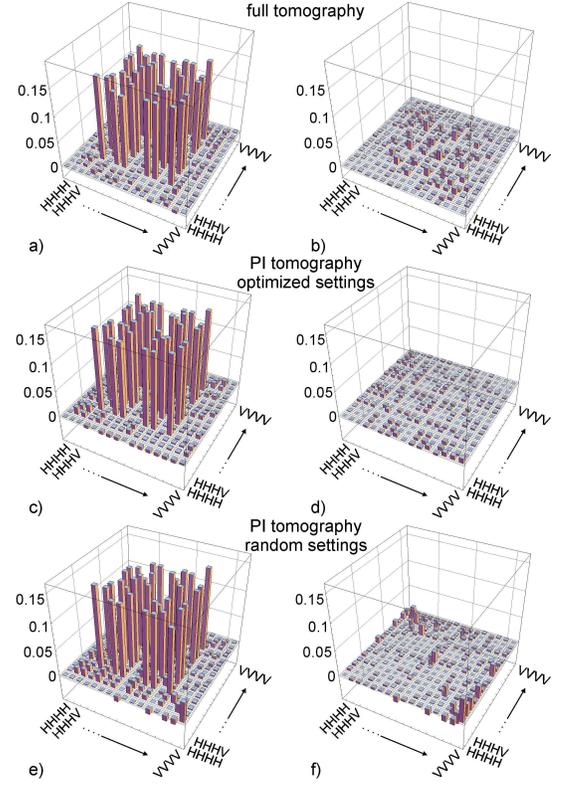}
\par\end{centering}

\caption{\label{fig:dmat}(Color online) (a) The real and (b) imaginary parts of the density matrix
coming from full tomography. (c),(d) The same for permutationally invariant tomography with optimized and (e),(f) random measurement directions, respectively.}
\end{figure}

\textbf{\emph{Experimental results.}} We demonstrate the method and the benefits of our algorithm for PI
tomography for a $4$-qubit symmetric Dicke state with two
excitations $\vert D_{4}^{(2)}\rangle$.
 First, we optimize the $\vec{a}_j$'s and the $c_j^{(k,l,m)}$'s for
 $\varrho_0=\openone/16$ and only for the uncertainty of full four-qubit
correlation terms, which means that when
computing $\mathcal{E}_{\rm total},$ we
 carry out the summation in Eq.~\eqref{eq:ET} only for the terms
with $n=0.$
With simple numerical optimization,
we were looking for the set of $A_{j}$ basis matrices that minimize the
uncertainty of the full correlation terms. Then, we also looked for
the basis matrices that minimize the sum of the squared error of all
the Bloch vector elements and considered also density matrices different from white noise,
such as a pure Dicke state mixed with noise.
We find that the gain in terms of decreasing the uncertainties is negligible in our
case and that it is sufficient to optimize for $\varrho_0=\openone/16$ and
for the full correlation terms. To demonstrate the benefits of the optimization of
the measurement directions, we also compare the results with those obtained with
randomly distributed basis
matrices.

The Dicke state was observed in a photonic system.
Essentially, four photons emitted by the second-order collinear type-II
spontaneous parametric down-conversion process were symmetrically distributed into four spatial modes.
Upon detection of one photon from each of the outputs, the state $\vert D_4^{(2)} \rangle$ is observed. Polarization analysis in each mode is used to characterize the experimentally observed state.
We collected data for each setting for $5$ minutes, with an average count
rate of $410$ per minute.
The experimental setup
has been described in detail in Refs.~\cite{KS07,WK09}.

First, to check the applicability of the PI tomography,
we apply our tools described above requiring only the measurement of
the three settings, $X^{\otimes 4}, Y^{\otimes 4}$ and $Z^{\otimes 4}.$
We determine the expectation value of the projector to the symmetric
subspace, yielding $\langle P_{\rm s} \rangle\ge 0.905 \pm 0.015.$
Based on Observation 2, we obtain $F(\varrho,\varrho_{PI})\ge 0.819\pm0.028.$
These results
show that the state is close
to be PI and has a large overlap with the symmetric subspace.
Thus, it makes sense to apply PI tomography.

For PI tomography of a four-qubit system, the measurement of
$15$ settings is needed.
 We used Eq.~\eqref{eq:XYZ2} to
obtain the Bloch vector elements from the experimentally measured
quantities.
This way, we could obtain all the $34$
symmetric correlations of the form $(X^{\otimes k}\otimes Y^{\otimes l}\otimes Z^{\otimes m}\otimes\openone^{\otimes n})_{{\rm PI}}$.
In Fig.~\ref{fig:Error-of-the}, we give the values of the correlations
for optimized and for randomly
chosen measurement directions, compared to the results obtained from full tomography, which needed
$81$ measurement settings.  As can be seen in Fig.~\ref{fig:Error-of-the},
the uncertainty for the optimized settings is considerably smaller than the one
for the randomly chosen settings. Moreover, the results from the optimized settings
fit very well the results of the full tomography. In Fig.~\ref{fig:dmat}, we compare the density matrices obtained
from full tomography [Fig.~\ref{fig:dmat}(a)], from PI tomography for optimized [Fig.~\ref{fig:dmat}(b)] and for random measurement directions [Fig.~\ref{fig:dmat}(c)].
Because of noise, the fidelity of the result of the full tomography with respect to $\vert D_4^{(2)}\rangle$ is
$0.873\pm0.005,$ which is similar to the fidelity of the results of the PI
tomography with optimized settings, $0.852\pm0.009$ \cite{expval}. In contrast, for the method using random
measurement directions, the fidelity is  $0.814\pm0.059,$ for which the uncertainty is the largest compared to all previous fidelity values. Finally, we also computed the fidelity of the results with respect to the PI density matrix obtained from
full tomography \cite{fitting}.
The results of the PI tomography with optimized settings shows a good agreement
with full tomography, the fidelity is $0.947,$ which is quite close to
the fidelity between the results of full tomography and
its PI part, $0.964.$
On the other hand, for the PI tomography with
random settings the corresponding fidelity is much lower, $0.880.$
Overall, the PI tomography shows a good agreement with the full tomography for this particular experiment.
However, a reasonable choice of measurement directions is needed to
obtain uncertainties in the reconstructed Bloch vector elements similar
to the ones from full tomography.

Finally, let us comment on how our method can be extended to lager systems.
Permutationally invariant operators can be represented efficiently on a digital computer in the basis of
$(X^{\otimes k}\otimes Y^{\otimes l}\otimes Z^{\otimes m}\otimes\openone^{\otimes n})_{{\rm PI}}$
operators. We determined the optimal $A_j$ operators for PI tomography
for systems with $N=6,8,...,14$ qubits.
To have the same maximum uncertainty of the Bloch vector elements as for the
$N=4$ case, one has to increase  the counts per setting by less than $50\%$
\citeappendix{(see the Appendix).}
\citepreprint{\cite{preprint}.}

In summary, we presented a scalable method for permutationally invariant tomography,
which can be used in place of full state tomography in experiments
that aim at preparing permutationally invariant many-qubit states.
For our approach, the same operator has to be measured on all qubits,
which is a clear advantage in some experiments.
We showed how to choose
the measurements such that the uncertainty in the reconstructed density
matrix is the smallest possible. This paves the way of characterizing
permutationally invariant states of many qubits in various physical systems.
Moreover, this work also shows that, given some knowledge or justifiable assumptions,
there is a way to obtain scalable state tomography for multiqubit entangled states.
\begin{acknowledgments}
We thank D.~Hayes and N.~Kiesel for discussions. We thank the
the Spanish MEC
(Consolider-Ingenio 2010 project ''QOIT'', Project No. FIS2009-12773-C02-02),
the Basque Government (Project No. IT4720-10), the ERC StG GEDENTQOPT,
the DFG-Cluster of Excellence MAP, the EU projects
QAP, Q-Essence and CORNER, and the DAAD/MNISW for support. W.W. and C.S. thank
the QCCC of the Elite Network of Bavaria for support.
\end{acknowledgments}


\eject

\renewcommand{\thefigure}{S\arabic{figure}}
\renewcommand{\thetable}{S\arabic{table}}
\renewcommand{\theequation}{S\arabic{equation}}
\setcounter{figure}{0}
\setcounter{table}{0}
\setcounter{equation}{0}

\subsection{Supplementary Material}

The supplement contains some derivations to help to understand the details of
the proofs of the main text. It also contains some additional experimental results.

\textbf{\emph{Proof of that we have to measure the same operator on all qubits.}}
From the proof of Observation 1,
we know that at least $\mathcal{D}_N$ measurements are needed to get the expectation values of all
 the $\mathcal{D}_N$ independent symmetric full
$N$-particle correlations.
What if we measure $\mathcal{D}_N$ settings, but several of them are not $\{A_j,A_j,...,A_j\}$-type, but $\{A_j^{(1)},A_j^{(2)},...,A_j^{(N)}\}$-type, i.e., we do not measure the same operator on all qubits? Each setting makes
it possible to get a single operator containing full $N$-qubit correlations. Let us denote this operator by $M_k$ for $k = 1,2,..., \mathcal{D}_N.$
Then, we know the expectation value of any operator of the space defined by the $M_k$ operators.
However, not all $M_k$'s are permutationally invariant. Thus, the size of the PI subspace of the space of the $M_k$
operators is less than $\mathcal{D}_N.$
We do not have $\mathcal{D}_N$
linearly independent symmetric operators in this space.
Thus, $\mathcal{D}_N$ measurement settings are sufficient to measure $\varrho_{\rm PI}$ only if we
have settings of the type  $\{A_j,A_j,...,A_j\}.$

\textbf{\emph{Derivation of Eq.~(7).}}
The eigen-decomposition of the correlation term is
 \begin{equation}
 (A_{j}^{\otimes(N-n)}\otimes\openone^{\otimes n})_{{\rm PI}}=\sum_k\Lambda_{j,n,k}
 \vert\Phi_{j,k}\rangle\langle\Phi_{j,k}\vert. \label{eigen}
\end{equation}
The individual counts
$N_C(A_j)_k$ follow a Poissonian distribution $f(n_c,\lambda_{j,k}),$
where $\lambda_{j,k}$ are the parameters of the Poissonian distributions and
$\sum_{k}\lambda_{j,k}=\lambda_j.$
The conditional variance, knowing that the total count is $N_C(A_j),$ is
\begin{equation}
\mathcal{E}^2[(A_{j}^{\otimes(N-n)}\otimes\openone^{\otimes n})_{{\rm PI}}\vert N_C(A_j)]=
\frac{[\Delta  (A_{j}^{\otimes(N-n)}\otimes\openone^{\otimes n})_{{\rm PI}}]^2}{N_C(A_j)}.
\end{equation}
After straightforward algebra, the variance is obtained as
\begin{eqnarray}
&&\mathcal{E}^2[(A_{j}^{\otimes(N-n)}\otimes\openone^{\otimes n})_{{\rm PI}}]\nonumber\\
&&\;\;\;\;\;\;\;=\sum_m f(m,\lambda_j) \mathcal{E}^2[(A_{j}^{\otimes(N-n)}\otimes\openone^{\otimes n})_{{\rm PI}}\vert N_C(A_j)=m]\nonumber\\
&&\;\;\;\;\;\;\;=\frac{[\Delta  (A_{j}^{\otimes(N-n)}\otimes\openone^{\otimes n})_{{\rm PI}}]^2}{\lambda_j-1}.
\end{eqnarray}
Similar results can be obtained through assuming Poissonian measurement statistics and Gaussian
error propagation [S1,S2].
If $\varrho_0=\openone/2^N,$ then $\Delta  (A_{j}^{\otimes(N-n)}\otimes\openone^{\otimes n})_{{\rm PI}}$ is independent from the choice of $A_j.$ By substituting $A_j=Z,$ straightforward calculations gives
\begin{equation}
\mathcal{E}^2[(A_{j}^{\otimes(N-n)}\otimes\openone^{\otimes n})_{{\rm PI}}]=
\frac{\binom{N}{n}^{-1}}{\lambda_j-1}.
\end{equation}

\begin{table*}
\caption{\label{tab1} Fidelities to the 4-qubit Dicke states.}
\begin{tabular}{l||c|c|c|c|c|c}
measurement&$\dicke{4}{0}$&$\dicke{4}{1}$&$\dicke{4}{2}$&$\dicke{4}{3}$&$\dicke{4}{4}$&$\Sigma$\\\hline\hline
full tomography& $-0.001\pm0.002$&$0.023\pm0.004$&$0.873\pm0.005$&$0.026\pm0.004$&$0.002\pm0.002$&$0.922$\\\hline
full tomography (max-like)&$0.001$&$0.021$&$0.869$&$0.023$&$0$&$0.914$\\\hline\hline
PI tomography &$-0.001\pm0.002$&$0.040\pm0.007$&$0.852\pm0.009$&$0.036\pm0.007$&$-0.002\pm0.002$&$0.925$ \\\hline
PI tomography (max-like)&$0.003$&$0.038$&$0.850$&$0.037$&$0$&$0.928$ \\\hline\hline
PI tomography (ran)&$0.000\pm0.002$&$0.055\pm0.027$&$0.814\pm0.059$&$0.023\pm0.027$&$0.001\pm0.002$&$0.893$\\\hline
PI tomography (ran,max-like)&$0.004$&$0.050$&$0.816$&$0.020$&$0.007$&$0.897$\\\hline
\end{tabular}
\end{table*}

\textbf{Obtaining the formula for $c_{j}^{(k,l,m)}$ for the smallest error.}
We look for $c_{j}^{(k,l,m)}$ for which the squared uncertainty given in
Eq.~(6) is the smallest.
In the following, we use the definition given in the main text for $\vec{c},$ $\vec{v},$ $V$ and $E.$
Thus, $V$ is matrix mapping a large space $\RR^l$ to a small space
$\RR^s$. Let $E$ be a non-singular diagonal matrix in the small space.
We have to
solve
\begin{eqnarray}
	\min_{\vec{c}} \| E \vec{c} \|^2\qquad
	&\mathrm{s.t.}\qquad&
	V \vec{c} = \vec{v},
\end{eqnarray}
where $\vert\vert \vec{a} \vert\vert$ is the  Euclidean norm of $\vec{a}.$
Using Lagrangian multipliers, we write down the condition for a minimum
fulfilling the constraints $V \vec{c} = \vec{v}$
\begin{eqnarray}
	\nabla_{\vec{c}} \big\{\vec{c}^T E^2 \vec{c} + \sum_{i=1}^s \lambda_i \big[(V \vec{c})_i - w_i \big]\big\}=0.
\end{eqnarray}
Hence, the condition for a local
(and, due to convexity, global) minimum is
\begin{eqnarray}
	\vec{c} = \frac12 E^{-2} V^T \vec{\lambda},
\end{eqnarray}
where $\lambda \in \RR^s$ is the vector of multipliers. In other
words, we have a minimum if and only if $\vec{c}\in\operatorname{range}
E^{-2} V^T$. Because the range of $V^T$ is an $s$-dimensional subspace
in $\RR^l$, there is a \emph{unique} $\vec{c}$ in that range such that $V\vec{c} = \vec{v}$.
A solution in a closed form can be obtained as
\begin{equation}\label{eqn:invert}
	c = E^{-2} V^T (V E^{-2} V^T)^{-1} \vec{v}.
\end{equation}
Simple calculation shows that the $V\vec{c} = \vec{v}$ condition holds
\begin{equation}
	V c = V E^{-2} V^T ( V E^{-2} V^T)^{-1} \vec{v} = \vec{v}.
\end{equation}

\textbf{\emph{Proof of Observation 2.}}  The eigenstates of
$\vec{J}^{2}=J_{x}^{2}+J_{y}^{2}+J_{z}^{2}$ are usually labelled
by $\vert j,m,\alpha\rangle,$ where $\vec{J}^{2}\vert j,m,\alpha\rangle= j(j+1)\vert j,m,\alpha\rangle,$
$J_z\vert j,m,\alpha\rangle= m \vert j,m,\alpha\rangle,$ and $\alpha$ is used to label the different eigenstates
having the same $j$ and $m$ [S3].
Let $P_{j,\alpha}$ denote the projector to the
subspace of a given $j$ and $\alpha.$ The
number of subspaces is denoted by $N_{\rm SS},$
and, for a given $N$, it can be calculated from group theory.
Moreover,  $P_{\rm s}\equiv P_{N/2,1}.$ Using this notation,
$\varrho_{{\rm PI}}=\sum_{j,\alpha}P_{j,\alpha}\varrho P_{j,\alpha}=(P_{{\rm s}}\varrho P_{{\rm s}})+\sum_{j<N/2,\alpha}(P_{j,\alpha}\varrho P_{j,\alpha}).$ In the basis of $\vec{J}^{2}$ eigenstates, $\varrho_{{\rm PI}}$
can be written as a block diagonal matrix
\begin{equation}
\varrho_{{\rm PI}}=\bigoplus_{j,\alpha}\left(\langle P_{j,\alpha}\rangle_{\varrho}\hat{\varrho}_{j,\alpha}\right),
\end{equation}
where $\hat{\varrho}_{j,\alpha}$ are density matrices of size $(2j+1)\times(2j+1).$ In another context,
\begin{equation}
\varrho_{{\rm PI}}=\sum_{j,\alpha}\langle P_{j,\alpha}\rangle_{\varrho}\varrho_{j,\alpha},\end{equation}
where $\varrho_{j,\alpha}=P_{j,\alpha}\varrho P_{j,\alpha}/{\rm Tr}(P_{j,\alpha}\varrho P_{j,\alpha}).$
Based on that, we obtain \begin{equation}
F(\varrho,\varrho_{j,\alpha})=\langle P_{j,\alpha}\rangle_{\varrho}.\label{eq:FP}\end{equation}
Then, due to the separate concavity of the fidelity, i.e., $F(\varrho,p_{1}\varrho_{1}+p_{2}\varrho_{2})\ge p_{1}F(\varrho,\varrho_{1})+p_{2}F(\varrho,\varrho_{2}),$ we obtain
$F(\varrho,\varrho_{{\rm {\rm PI}}})\ge\langle P_{{\rm s}}\rangle_{\varrho}F(\varrho,\varrho_{{\rm s}})+\sum_{j<N/2,\alpha}\langle P_{j,\alpha}\rangle_{\varrho}F(\varrho,\varrho_{j,\alpha}).$
 Substituting Eq.~\eqref{eq:FP} into this inequality, we obtain
 $ F(\varrho,\varrho_{{\rm {\rm PI}}})\ge\langle P_{{\rm s}}\rangle_{\varrho}^{2}+\sum_{j<N/2,\alpha}\langle P_{j,\alpha}\rangle_{\varrho}^{2}.$
 Using the fact that
 $\langle P_{\rm s}\rangle_{\varrho}+\sum_{j<N/2,\alpha}\langle P_{j,\alpha}\rangle_{\varrho}=1,$ we obtain
 \begin{equation}
 F(\varrho,\varrho_{{\rm PI}}) \ge\langle P_{{\rm s}}\rangle_{\varrho}^{2}
 +\frac{(1-\langle P_{{\rm s}}\rangle_{\varrho})^{2}}{ N_{\rm SS}-1  }.\label{vvv}
  \end{equation}
  In many practical situations, the state $\varrho$ is almost symmetric and $N$ is large.
  In such cases the second term in Eq.~(\ref{vvv}) is negligible.
 Thus, a somewhat weaker bound presented in
 Observation 2 can be used.

\begin{figure}
\includegraphics{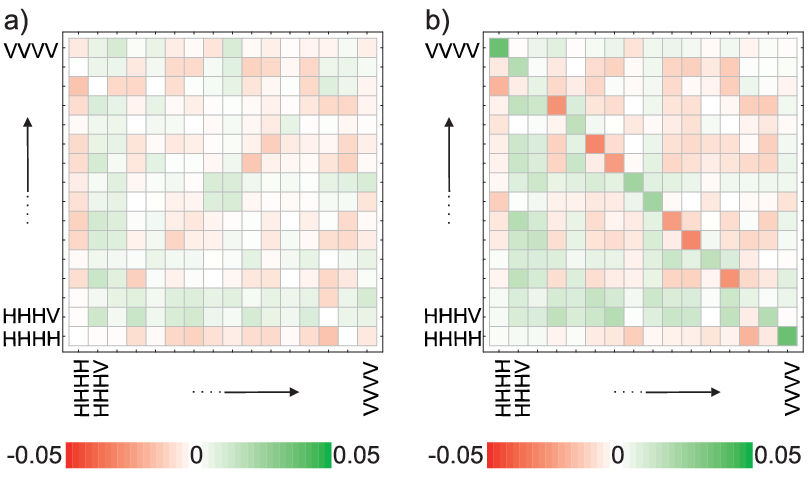}
\caption{\label{fig3} 
(a) The difference of the real part of the density matrices from optimized settings and the one of full tomography. (b) The difference of the density matrices from random settings and the one of full tomography. For the former, no clear structure is observed, whereas for the latter the largest difference is observed for the antidiagonal elements.}
\end{figure}

\textbf{Numerical optimization used to minimize $\mathcal{E}_{\rm total}.$} The measurement directions
minimizing $\mathcal{E}_{\rm total}$ can be obtained as follows. Let us represent the measurement directions
by three-dimensional vectors $\{\vec{a}_j\}_{j=1}^{\mathcal{D}_N}.$ The operators can be obtained as
$A_j=a_{j,x}X+a_{j,y}Y+a_{j,z}Z.$

First, we need an initial guess. This can come from a set of randomly chosen vectors representing the measurement directions.
One can also use the result of a minimization for some measure that characterizes how equally the vectors are
distributed. Such a measure is defined by
\begin{equation}
\mathcal{F}(\{v_j\})=\sum_{k,l}(\vec{v}_k\cdot \vec{v}_l)^{2m},
\end{equation}
where $\vec{v}_k$ represent the measurement directions and $\cdot$ is the scalar product and $m$ is an integer.
Such cost functions, called frame potentials, appear in the theory
of $t$-designs essentially for the  same purpose.

After we obtain the initial guess from such a procedure,
we start an optimization for decreasing $\mathcal{E}_{\rm total}.$
At each iteration of the method,
we change the measurement directions by rotating them with a small random angle around a randomly chosen axis.
If the change decreases $\mathcal{E}_{\rm total},$
then we keep the new measurement directions, while if it does not then we discard it. We repeat this procedure until
$\mathcal{E}_{\rm total}$ does not change significantly.

\textbf{Three-setting witness for estimating the fidelity} The three-setting witness for detecting genuine
multipartite entanglement in the vicinity of the Dicke state is [S4]
\begin{equation}
\mathcal{W}_{\rm D(4,2)}^{\rm (P3)}=2\cdot\openone + \tfrac{1}{6}(J_x^2+J_y^2-J_x^4-J_y^4)+\tfrac{31}{12}J_z^2-\tfrac{7}{12}J_z^4.
\end{equation}
For this witness we have [S4]
\begin{equation}
\mathcal{W}_{\rm D(4,2)}^{\rm (P3)}-3\mathcal{W}_{\rm D(4,2)}^{\rm (P)}\ge 0,
\end{equation}
where the projector witness is defined as
\begin{equation}
\mathcal{W}_{\rm D(4,2)}^{\rm (P)}=\tfrac{2}{3}\cdot\openone-\vert D_4^{(2)}\rangle \langle D_4^{(2)} \vert.
\end{equation}
Hence, the fidelity with respect to the state $\vert D_4^{(2)}\rangle $ is bounded from below as [S4]
\begin{equation}
F_{\rm D(4,2)} \ge \tfrac{2}{3}-\tfrac{1}{3} \langle \mathcal{W}_{\rm D(4,2)}^{\rm (P3)} \rangle.
\end{equation}

\textbf{Fidelities with respect to the four-qubit Dicke states.} In Table~\ref{tab1} we summarize the results for full tomography (full) and for permutationally invariant tomography (PI) for random (ran) and optimized (opt) directions.
To obtain a physical density matrix with non-negative eigenvalues we perform a maximum-likelihood fit (max-like) of the measured data.
In Fig.~\ref{fig3}, the differences between the density matrix obtained from full tomography and the ones obtained from permutationally invariant tomography can be seen.

\textbf{Efficient representation of permutationally invariant operators on a digital computer.} Every PI operator $O$ can be decomposed as
\begin{equation}
O=\sum_{k+l+m+n=N}c_{k,l,m,n}^{(O)}(X^{\otimes k}\otimes Y^{\otimes l}\otimes Z^{\otimes m}\otimes\openone^{\otimes n})_{{\rm PI}}.
\end{equation}
Such a decomposition for operators of the form $(A^{\otimes(N-n)}\otimes\openone^{\otimes n})_{{\rm PI}}$ with $A=a_{x}X+a_{y}Y+a_{z}Z$ is given by
\begin{equation}
\sum_{k,l,m}a_{x}^{k}a_{y}^{l}a_{z}^{m}\frac{(k+l+m)!}{k!l!m!}(X^{\otimes k}\otimes Y^{\otimes l}\otimes Z^{\otimes m}\otimes\openone^{\otimes n})_{{\rm PI}},
\end{equation}
where the summation is carried out such that $k+l+m+n=N.$

\begin{figure}
\includegraphics[width=5.2cm]{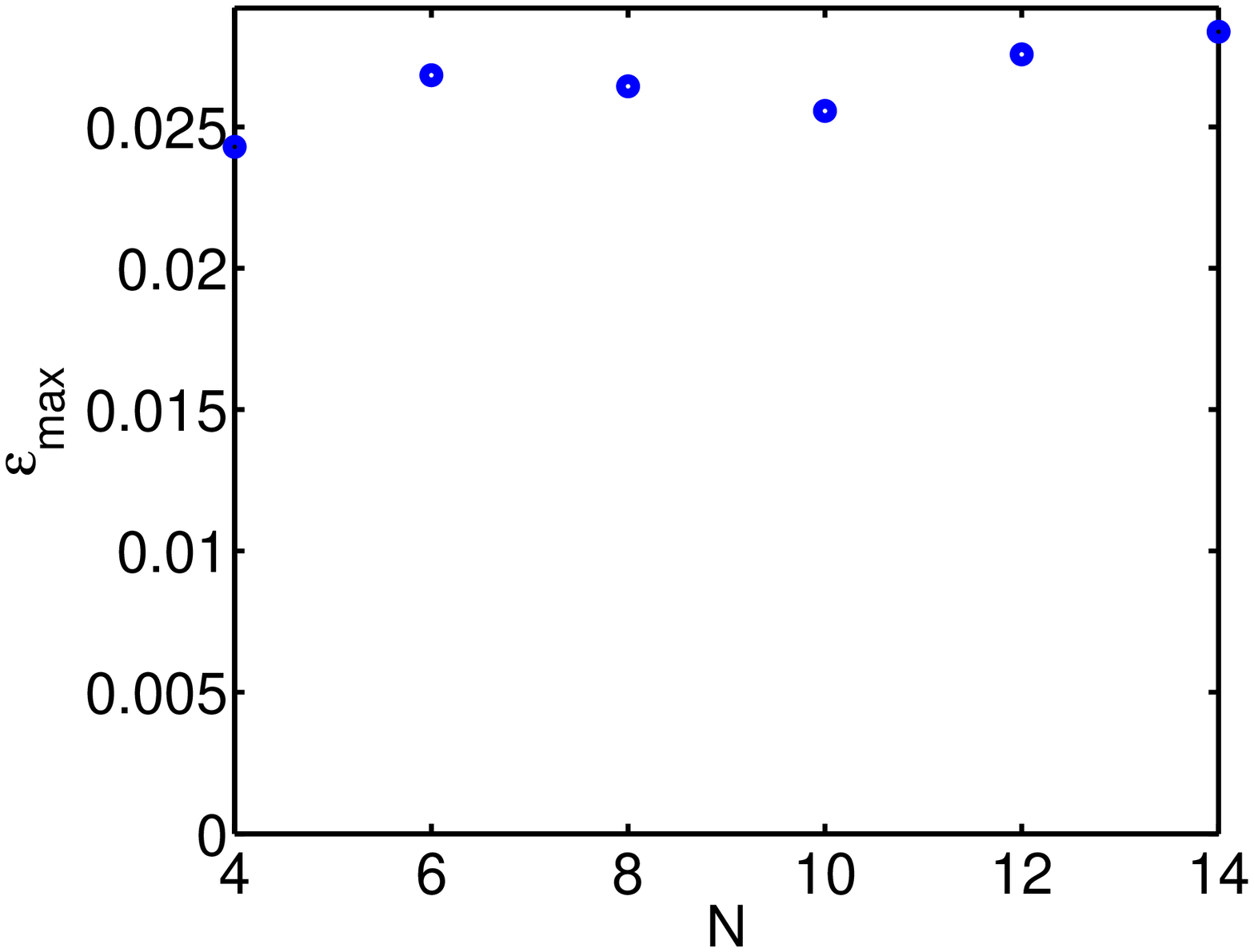}
\caption{\label{fig4} 
The maximum uncertainty of the Bloch vector elements defined in Eq.~(\ref{ms}) for the optimal measurement settings as a function of the number of qubits, $N,$ for $N=4,6,8,10,12$ and $14.$}
\end{figure}

\textbf{Results for larger systems.} We determined the optimal $A_j$ for PI tomography for $N=4,6,...,14.$ In Fig.~\ref{fig4}, we plot the maximal uncertainty of the Bloch vector elements
\begin{equation}\epsilon_{\max}=\max_{k,l,m,n}\mathcal{E}[(X^{\otimes k}\otimes Y^{\otimes l}\otimes
	Z^{\otimes m}\otimes\openone^{\otimes n})_{{\rm PI}}]\label{ms}\end{equation} for the total count realized in the experiment
$\lambda_j=\lambda=2050$ as a function of $N,$ when the state of the system is $\varrho_0=\openone/2^N.$
It increases slowly with $N.$ Thus, for large $N$ the number of counts per measurement setting
does not have to increase very much in order to keep
the maximal uncertainty of the Bloch vector elements the same as for the $N=4$ case.
In particular, for $N=14,$ a total count of $2797$ per setting yields the same maximal uncertainty
as we had for the $N=4$ case.

An upper bound on the uncertainty of PI tomography for $\varrho_0$ different from the white noise
 can be obtained by using
$[\Delta (A_{j}^{\otimes(N-n)}\otimes\openone^{\otimes n})_{{\rm PI}}]^2_{\varrho_0}=1$
for error calculations. According to numerics, for optimal $A_j$ for $N=4,6,...,14,$
$\epsilon_{\max}$ remains the same as in the case of white noise,
since for the full correlation terms
with $n=0$ the upper bound equals the value for white noise, and the full correlations terms
contribute to the noise of the Bloch vector elements with the largest uncertainty.
Thus, the
total count per setting will not increase more with the number of qubits even for states different
from the completely mixed state.

The operators that give a bound on $\langle P_{\rm s}\rangle$ with three settings for $N=6$ and $8$ are the following
\begin{eqnarray}
P_{\rm s}^{(6)}&\ge& \tfrac{2}{225}(Q_2+J_z^2)-\tfrac{1}{90}(Q_4+J_z^4)+\tfrac{1}{450}(Q_6+J_z^6) ,\nonumber\\
P_{\rm s}^{(8)}&\ge& -0.001616Q_2+0.002200Q_4-0.0006286Q_6\nonumber\\
                     &+&0.00004490Q_8+0.003265J_z^2-0.004444J_z^4\nonumber\\
                     &+&0.001270J_z^6-0.00009070J_z^8,\label{Ps}
\end{eqnarray}
where $Q_n=J_x^n+J_y^n.$
They were determined using semi-definite programming, with a method similar to one used for obtaining three-setting witnesses in Ref.~[S4]. They have an expectation value $+1$ for the Dicke states $\vert D_{6}^{(3)}\rangle$
and $\vert D_{8}^{(4)}\rangle,$ respectively. Moreover, their expectation value give the highest possible lower bound on
$\langle P_{\rm s} \rangle$ for states of the form
\begin{eqnarray}
\varrho_{\rm noisy}(p)=p\frac{\openone}{2^N}+(1-p)\vert D_{N}^{(N/2)}\rangle\langle D_{N}^{(N/2)}\vert
\end{eqnarray}
among the operators that are constructed as a linear combination of the operators $J_l^n.$
The validity of the relations in Eq.~(\ref{Ps}) can easily be checked by direct calculation.

\textbf{Bounding the differences between elements of $\varrho$ and $\varrho_{\rm PI}$ based on the fidelity.}
For any pure state $\vert\Psi\rangle,$ it is possible to bound the difference between
$\vert\langle\Psi\vert\varrho_{\rm PI}\vert\Psi\rangle\vert$ and $\vert\langle\Psi\vert\varrho\vert\Psi\rangle\vert$ as
\begin{equation}
\vert\langle\Psi\vert\varrho\vert\Psi\rangle-\langle\Psi|\varrho_{{\rm PI}}
|\Psi\rangle\vert\le\sqrt{1-F(\varrho,\varrho_{{\rm PI}})}.\label{ppp}
\end{equation}
 Thus, if the fidelity is close to $1,$ then $\langle\Psi\vert\varrho_{{\rm }}\vert\Psi\rangle\approx\langle\Psi\vert\varrho_{{\rm {\rm PI}}}\vert\Psi\rangle,$ even if $\vert\Psi\rangle$ is non-symmetric. If $\vert\Psi\rangle$ is an element of the product basis, e.g.,
 $\ket{0011},$ then Eq.~(\ref{ppp}) is a bound on the difference between the corresponding diagonal elements of
 $\varrho$ and $\varrho_{\rm PI}.$

Eq.~(\ref{ppp}) can be proved as follows: There is a well-known relation between the trace norm and the fidelity [S5]
\begin{equation}
\frac{1}{2} \vert\vert\varrho - \varrho_{\rm PI}\vert\vert_{\rm tr} \le \sqrt{1-F(\varrho,\varrho_{{\rm PI}})}.\label{ppp1}
\end{equation}
Moreover, for a projector $P$ and density matrices $\varrho_{k}$ we have [S6]
\begin{equation}
\vert {\rm Tr}(P\varrho_1)-   {\rm Tr}(P\varrho_{2})   \vert\le\frac{1}{2} \vert\vert\varrho_1 - \varrho_{2}\vert\vert_{\rm tr}.\label{ppp2}
\end{equation}
Combining Eq.~(\ref{ppp1}) and Eq.~(\ref{ppp2}), leads to Eq.~(\ref{ppp}).

\vskip0.6cm
{\bf References}

\begin{enumerate}

\item[{[}S1{]}] C.~Schmid, Ph.D. Thesis, Ludwig-Maximillian-Universit\"at,
Munich, Germany, 2008; D.F.V.~James \emph{et. al, } Phys. Rev. A \textbf{64},
052312 (2001).

\item[{[}S2{]}] B.~Jungnitsch \emph{et. al}, Phys. Rev. Lett.  \textbf{104}, 210401 (2010).

\item[{[}S3{]}] See, for example, J.I. Cirac  \emph{et. al}, Phys. Rev. Lett.  \textbf{82}, 4344 (1999).

\item[{[}S4{]}] G.~T\'oth \emph{et al.}, New J. Phys. \textbf{11}, 083002 (2009).

\item[{[}S5{]}] J.A.~Miszczak \emph{et al.},  J. Quant. Inf. Comp. {\bf 9}, 0103 (2009).

\item[{[}S6{]}] See Eqs.~(9.18) and (9.22) of M.A. Nielsen and I.L. Chuang,
{\it Quantum computation and quantum information} (Cambridge University
Press, Cambridge, 2000).

\end{enumerate}


\begin{thebibliography}{25}
\bibitem{expval} Expectation values are obtained directly from the measured data, rather than from $\rho_{\rm PI}.$
\bibitem{fitting} Values without error are deduced from fitted matrices obtained via maximum likelihood estimation \cite{S09}.
\end{thebibliography}
\end{document}